\documentclass[12pt,a4]{article}
\usepackage{psfig,epsfig,graphicx}
\usepackage{amsmath}
\usepackage{amssymb}

\begin{document}

\title{Kinetics calculation on the  shear viscosity in hot QED at finite density}
\author{Liu Hui \thanks{liuhui@iopp.ccnu.edu.cn} \ \ Hou Defu \thanks{hdf@iopp.ccnu.edu.cn}
 \ \ Li Jiarong \thanks{ljr@iopp.ccnu.edu.cn} \\[0.5cm] {\it\small Institute of Particle Physics, Central China Normal University,}\\ {\it
\small Wuhan(430079), P.R.China}}
\date{}
\maketitle
\begin{abstract}
The shear viscosity of QED plasma at finite temperature and
density is calculated  by solving Boltzmann equation
with variational approach. The result shows the small chemical
potential enhances the viscosity in leading-log order by adding a
chemical potential quadratic term  to the viscosity for the pure temperature environment.
\end{abstract}

\section{Introduction}
A novel  state of matter, strongly interacting  quark-gluon
plasma(sQGP) is claimed to be found at Relativistic Heavy Ion
Collider at Brookhaven National Laboratory\cite{Shuryak}. The
measured $v_2$  was found to reach the hydrodynamic limit of an
almost perfect fluid with very small viscosity at low transverse
momentum region. It is desirable  to explain this near-perfect
fluid behavior of sQGP from the theoretical points of
view\cite{Son}.

In principle, there are two approaches to calculate transport
coefficients. One is using the Kubo formulae\cite{Hosoya} within
the thermal field theory, with which people evaluated the shear
viscosity via resumming an infinite series of ladder diagrams
\cite{Jeon,Carrington2,wang2}.  The alternative framework is the
kinetics theory\cite{Baym,Arnold1,Arnold2,Heiselberg}. Although
the transport equations are hard to solve,  the relaxation time
approximation(RTA) and variational calculus are two popular
methods to obtain the transport coefficients. In RTA, people use
classic kinetic formulae, but involving the relativistic and
quantum effects, to estimate the shear
viscosity\cite{Danielewicz,Thoma2,Defu}. Arnold, Moore and
Yaffe\cite{Arnold1,Arnold2} have studied the leading-log
contribution as well as the full leading order contribution of
various transport coefficients of the QCD-like theory at high
temperature by solving the Boltzmann equation with variational
approach. The results in the two frameworks are coincident in
leading-log order except for some factor differences. Some
publications also demonstrated that the diagrammatic expansion of
Kubo formula was actually equivalent to the kinetics calculation
from the linearized Boltzmann equation if all the possible ladder
diagrams were resummed in scalar field\cite{Carrington3} and in
pure gauge theory\cite{Basagoti,Defu2}. In addition, one should
pay attention to the consistency of the Ward identity with the
ladder resummation\cite{Aarts} in gauge theory.

However, most works listed above concentrated on the high
temperature but vanishing chemical potential except
Ref.\cite{Defu}. While actually the net baryon number in the
central fire ball of heavy-ion collision is not zero rigidly
though small\cite{Braun}. It makes sense to involve this density
effect by introducing a chemical potential $\mu$, which is much
smaller than the temperature, to study how it affects the shear
viscosity of the plasma.

In this paper, we shall try to solve the Boltzmann equation by the
variational method at high temperature with finite density in QED,
following the scheme in Ref.\cite{Arnold1} for high temperature
and zero chemical potential.
  QED is a good toy
model for the non-Abelian gauge QCD yet simpler in computation. We
found the shear viscous coefficient is proportional to
$T^3e^4/(\ln\frac{1}{e})$ and modified by a small factor of
$(1+0.13\mu^2/T^2)$.

The paper is arranged as following: in the second section, we will
review the sketch of solving Boltzmann equation by variational
method in the kinetics of transport theory and define the shear
viscosity in this framework. The associated collision processes on
the right hand side of Boltzmann equation will be calculated in
 section 3. And in the fourth section, we use the variational
method to obtain the shear viscosity. Section 5 is conclusion and
outlook.

We  use the notation $P=(p_0,\mathbf{p})$ and $p\equiv|\mathbf{p}|$.
 The  momentum denoted by a capital letter is the four-component
momentum and the lowercase with bold face denotes the
three-component momentum.

\section{Boltzmann equation and viscosity}

Considering a system which slightly deviates from the equilibrium
state by a small velocity gradient, one can describe it with the
one particle distribution which is satisfied the Boltzmann
equation
\begin{equation}\label{Boltzmann}
\left(\frac{\partial}{\partial t}+\mathbf{v_p}\cdot
\frac{\partial}{\partial
\mathbf{x}}+\mathbf{F}\cdot\frac{\partial}{\partial\mathbf{p}}\right)f(\mathbf{p};\mathbf{x},t)=-C[f],
\end{equation}
where $\mathbf{v_p}=\hat{\mathbf{p}}\equiv\mathbf{p}/p$ and
$\mathbf{F}$ is the external force. In the case of shear
viscosity, the external field is irrelevant and the time
derivation on the left hand side may be dropped out due to its
higher order contribution in spacial gradients\cite{Arnold1}. The
right hand side of equation(\ref{Boltzmann}) is the collision term
which takes the usual form of
\begin{eqnarray}
C[f](\mathbf{p})&=&\frac{1}{2}\int_{\mathbf{p'},\mathbf{k},\mathbf{k'}}|\mathcal{M}(p,k;p',k')|^2(2\pi)^4\delta^{(4)}(P+K-P'-K')\\[0.3cm]
&\times&\left\{f(\mathbf{p})f(\mathbf{k})[1\pm f(\mathbf{p})][1\pm
f(\mathbf{k})]-f(\mathbf{p'})f(\mathbf{k'})[1\pm
f(\mathbf{p'})][1\pm f(\mathbf{k'})]\right\}\nonumber,
\end{eqnarray}
if only $2\rightarrow 2$ elastic collisions are involved. Here
$\mathbf{p},\mathbf{k},\mathbf{p'}$ and $\mathbf{k'}$ denote the
momenta of the incoming and outgoing particles respectively. The
momentum space integration $\int_{\mathbf{p}}$ is a shorthand for
$\int\frac{d^3\mathbf{p}}{(2\pi)^3 2p_0}$,  and $|\mathcal{M}|^2$
is the two-body scattering amplitude. The $1\pm f$ factor is
 the final state statistical weight for boson with the upper sign
 and for fermion with the down sign.

Following Ref.\cite{Arnold1}, we expand the distribution function
in the near-equilibrium state and obtain the linearized Boltzmann
equation:
\begin{equation}
  S_{ij}(\mathbf{p})=\mathcal{C}\chi_{ij}(\mathbf{p}),
\end{equation}
where $\mathcal{C}$ is the linearized collision operator. By
defining the inner product in function space and variating the
trial function $\chi_{ij}(\mathbf{p})$, one can obtain the shear
viscous coefficient
\begin{equation}
\eta=\frac{2}{15}Q_{max}
\end{equation}
 where
\begin{eqnarray}
&&Q_{max}=\frac{1}{2}(\chi_{ij},\mathcal{C} \chi_{ij})
|_{\chi=\chi_{max}}=\frac{1}{2}(\chi_{ij},S_{ij})|_{\chi=\chi_{max}},\\
&&\chi_{ij}(\mathbf{p})=I_{ij}(\hat{\mathbf{
p}})\chi(p)=\sqrt{\frac{3}{2}}(\hat p_i\hat
p_j-\frac{1}{3}\delta_{ij})\chi(p), \\
&&(\chi_{ij},S_{ij})=-\beta^2\sum\limits_a \int_{\mathbf{p}}\
\mathbf{p}\ f^a_0(\mathbf{p})[1\pm f^a_0(\mathbf{p})] \chi^a
(p)\label{leftside}
\end{eqnarray}
and the collision term at the right hand side of Boltzmann
equation is
\begin{eqnarray}\label{Boltamann2}
(\chi_{ij},\mathcal{C}\chi_{ij})&=&\frac{\beta^2}{8}\int_{\mathbf{p},\mathbf{k},\mathbf{k},\mathbf{k'}}
\sum\limits_{abcd}|\mathcal{M}^{ab}_{cd}|^2\ (2\pi)^4\
\delta^{(4)}(P+K-P'-K')\nonumber\\[0.3cm]
&\times& f^a_0(\mathbf{p})f^b_0(\mathbf{k})[1\pm
f^c_0(\mathbf{p'})][1\pm f^d_0(\mathbf{k'})]\nonumber \\[0.3cm]
&\times&\left
[\chi_{ij}^a(\mathbf{p})+\chi_{ij}^b(\mathbf{k})-\chi_{ij}^c(\mathbf{p'})-\chi_{ij}^d(\mathbf{k'})\right
]^2.
\end{eqnarray}
where $a,b,c$ and $d$ are for species of particles.

In the above definitions, we adopted the formalisms developed by
Arnold, Moore and Yaffe\cite{Arnold1} with the only differences in
the distribution functions which involved the chemical potential
in the initial and the final states. Another notation one should
notice is the sum in front of the matrix element which means all
possible collision processes relevant to the leading-log
contribution are involved and properly treated without double
counting or multi-counting.

\section{Collision terms}

In QED, all possible reactions can be classified as two
categories: processes of exchanging a boson(Fig.1(a)) and
processes of exchanging a fermion(Fig.1(b) and (c)), in which the
later includes the pair production and the Compton scattering
processes. Notice that the s-channel scattering is omitted because
it is infrared finite thus does not contribute to the leading-log
result.
\begin{figure}
 \begin{center}
   \resizebox{12cm}{!}{\includegraphics{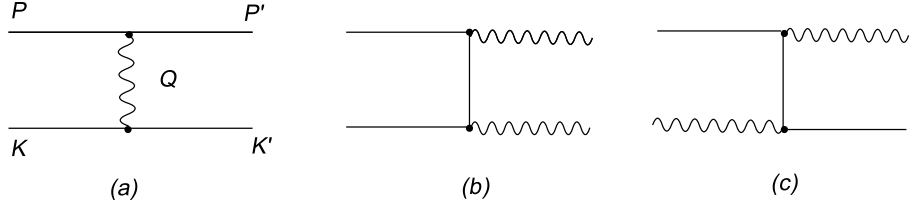}}
 \end{center}
   \caption{The possible processes which contribute to the leading-log in the collision term in QED plasma. The solid line is for electron and the wiggly line is for photon.}
\end{figure}


Before going into the next step of calculation, we should specify
some important approximations and definitions.
\begin{itemize}

\item In our discussion, we adopt the hard forward scattering
approximation, namely the  momentum transfer $q\sim eT$ which is
small for all the time since it is sufficient to compute the
leading-log viscosity. So we neglect all the differences between
the distribution functions such as $f(p)$ and $f(p')$. The fermion
mass is also omitted in this case, for it is in order of $eT$
which is much smaller than the hard scale $T$. Thus the kinematics
of the two-body collision gives
\begin{equation}
\cos\theta_{pk}=1+(1-\cos^2\theta)(1-\cos\phi)
\end{equation}
where $\theta_{pk}$ is the angle between $\mathbf{p}$ and
$\mathbf{k}$. $\theta$ is the angle between $\mathbf{p}$ and
$\mathbf{q}$, and the angle between $\mathbf{k}$ and $\mathbf{q}$
as well, since they are approximately equal in the forward
scattering. $\phi$ is the angle between the
$\mathbf{p}$-$\mathbf{q}$ plane and $\mathbf{p}$-$\mathbf{q}$
plane.

\item Due to the energy-momentum conservation, only three of the
four momenta of incoming and outgoing particles are independent.
If we properly label the particles coming from the same vertex
with the similar momentum symbols as shown in the Figure 1(a), for
example $P$ and $P'$, all the three Mandelstam variables can be
defined as $s=(P+K)^2$, $t=(P-P')^2$ and $u=(P-K')^2$.

\item As to the infrared divergence, the two categories of
collisions behave different. When the momentum transfer $q\equiv
|\mathbf{p}-\mathbf{p'}|$ goes to zero in the forward scattering,
 one finds the infrared singularity in
the fermion-exchange process is  logarithmical while in the boson
exchange process it is quadratic. Fortunately that is not so bad
for the latter case because if carefully considering the
$[\chi^a+\chi^b-\chi^c-\chi^d]^2$ term one may find a small $q^2$
emerges which softens the quadratic divergence into a
logarithmical one. Since now all the collision integrations are
logarithmically divergent, the limit cut-offs play important roles
in our calculation. For transport coefficients like shear
viscosity, these integrations are dominated by the hard scale $T$
of the system which can be chosen as the ultraviolet cut-off. As
to the infrared limit, the hard thermal loop self-energy scale
$eT$ is sufficient\cite{Arnold1}. Even in the finite density case,
the small chemical potential only modifies the infrared cut-off by
adding a factor like $e\mu$ behind $eT$, which does not contribute
to the leading-log order $\ln{1/e}$ since we assume the chemical
potential is much small than the typical momentum scale $T$, i.e.
$\mu\lesssim eT \ll T$. Therefore we will not carefully treat the
$dq$ integration and just adopt $T$ and $eT$ as the upper and down
limits respectively.
\end{itemize}

Now let us continue our calculation.
$\delta^3(\mathbf{p}+\mathbf{k}-\mathbf{p'}-\mathbf{k'})$ in the
integrand of equation(\ref{Boltamann2}) helps to perform the
$\mathbf{k'}$ integration, yet to the $\delta$ function of energy
conservation, one may introduce a dummy integration variable
$\omega$\cite{Baym}
\begin{equation}
\delta(p+k-p'-k')=\int^\infty_{-\infty}d\omega
\delta(\omega+p-p')\delta(\omega-k+k').
\end{equation}
With this trick we can integrate over the angles and the remaining
integrals are
\begin{eqnarray}\label{Boltamann3}
(\chi_{ij},\mathcal{C}\chi_{ij})&=&\frac{\beta^3}{(4\pi)^6}\int_0^\infty
dq \int_{-q}^q d\omega \int_0^\infty dp \int_0^\infty dk
\int_0^{2\pi}d\phi
\sum\limits_{abcd}|\mathcal{M}^{ab}_{cd}|^2\nonumber\\[0.3cm]
&\times& f^a_0(\mathbf{p})f^b_0(\mathbf{k})[1\pm
f^c_0(\mathbf{p})][1\pm f^d_0(\mathbf{k})]\nonumber \\[0.3cm]
&\times&\left
[\chi_{ij}^a(\mathbf{p})+\chi_{ij}^b(\mathbf{k})-\chi_{ij}^c(\mathbf{p'})-\chi_{ij}^d(\mathbf{k'})\right
]^2.
\end{eqnarray}
with $p'=p+\omega$ and $k'=k-\omega$. For the sake of convenience,
we adopt $f(p)$ as the fermion distribution and $b(p)$ for boson
function in the equilibrium state  in the following calculation.

\subsection{Boson-exchange Processes}
Unlike the pure temperature case, the system with finite chemical
potential requires more careful treatment to distinguish the
different species of fermions with different distribution
functions. For the boson-exchange process, Bhabha scattering
$e^+e^-\rightarrow e^+e^-$ and M{\o}ller scattering
$e^-e^-\rightarrow e^-e^-$ or $e^+e^+\rightarrow e^+e^+$ have been
involved. Since the s-channel has been omitted since it does not
contribute to the leading-log order, the distribution functions in
the Boltzmann equation for both scatterings are
\begin{eqnarray}
\mbox{Bhabha scattering}:&& 2\bar f(p)\bar f(k)[1-\bar f(p)][1-\bar f(k)]\label{df1}\nonumber\\
&&\ \ \ \ +2f(p)f(k)[1-f(p)][1-f(k)],\\[0.3cm]
\mbox{M{\o}ller scattering}&:& 4\ \bar f(p) f(k)[1-\bar
f(p)][1-f(k)]
\end{eqnarray}
where the extra factor of 4 in the M{\o}ller scattering process is
from the sum over the initial and final states, and the factor of
2 in the Bhabha scattering comes from the t-channel and u-channel
contributions respectively. $\bar f(p)=[e^{\beta (p+\mu)}+1]^{-1}$
is the distribution function for the positron and $ f(p)=[e^{\beta
(p-\mu)}+1]^{-1}$ is the distribution function for the electron.

In the forward scattering, one can easily check $s\approx-u$, thus
 the matrix element for t-channel is
\begin{equation}\label{me1}
8e^4\frac{s^2+u^2}{t^2}\approx16e^4\frac{u^2}{t^2}=16e^4\frac{4p^2k^2}{q^2}(1-\cos\phi)^2.
\end{equation}
where the spins of initial and final states have been summed. As
to u-channel, the matrix element is identical with that of
t-channel as long as the momentum symbols are well defined.

In the case of ${q}$ being small and the particle species $a,b$
being identical to $c,d$(or $d,c$) respectively, one finds
\begin{eqnarray}
\chi^e_{ij}(\mathbf{p'})-\chi^e_{ij}(\mathbf{p})&=&\mathbf{q}\cdot\nabla\chi^e_{ij}(\mathbf{p})+\cdots\\
&\approx&\omega
I_{ij}(\hat{\mathbf{p}})\chi^e(p)'-\sqrt\frac{3}{2}(2\omega\hat{p_i}\hat{p_i}-q_i\hat{p_j}-q_j\hat{p_i})\frac{\chi^e(p)}{p}\nonumber
\end{eqnarray}
where $\chi^e(p)'=d\chi^e(p)/dp$. The square of the above equation
one obtains
\begin{equation}
[\chi^e(\mathbf{p'})-\chi^e(\mathbf{p'})]^2=\omega^2[\chi^e(p)']^2
+3\frac{q^2-\omega^2}{p^2}[\chi^e(p)]^2+\mathcal{O}(q^3).
\end{equation}
Here, the electron and positron have the same departure from the
equilibrium which is denoted by $\chi^e$.  One can prove that the
cross terms like
$[\chi^e_{ij}(\mathbf{p'})-\chi^e_{ij}(\mathbf{p})]\cdot[\chi^e_{ij}(\mathbf{k'})-\chi^e_{ij}(\mathbf{k})]$
vanishes when carrying out $d\omega$ and $d\phi$ integration
 with the factor $(1-\cos\phi)^2$ coming from the matrix
element.

Combining the equations(\ref{Boltamann3}) and
(\ref{df1})-(\ref{me1}) and completing the $d\omega$ and $d\phi$
integration, we obtain the collision term for the boson-exchange
process
\begin{eqnarray}
(\chi_{ij},\mathcal{C}\chi_{ij})^{(a)}&=&\frac{\beta^3}{(2\pi)^3}\int_{eT}^T
\frac{dq}{q}\int_0^\infty dp \int_0^\infty  dk p^2 k^2
\{p^2[\chi^e(p)']^2+6[\chi^e(p)]^2\}\\[0.3cm]
&\times& \{f({p})f({k})[1-
f({p})][1-f({k})]+\bar f(p)\bar f(k)[1-\bar f(p)][1-\bar f(k)]\nonumber\\[0.3cm]
&&\ +\bar f(p)f(k)[1-\bar f(p)][1-f(k)]+f(p)\bar
f(k)[1-f(p)][1-\bar f(k)]\}\nonumber.
\end{eqnarray}
where we have replaced $k$ with $p$ in the $\chi$-functions and
placed an extra factor of 2 in front of the remaining integration.

Noticing that the k-integration can be done after expanding the
distribution functions in terms of $\mu/T$
\begin{eqnarray}
&&\int_0^\infty dk k^2 f(k)[1-f(k)]=T^3\left [\frac{\pi^2}{6}+\ln4\frac{\mu}{T}+\frac{\mu^2}{2T^2}+\mathcal{O}(\frac{\mu^3}{T^3})\right ],\\[0.3cm]
&&\int_0^\infty dk k^2 \bar f(k)[1-\bar
f(k)]=T^3\left[\frac{\pi^2}{6}-\ln4\frac{\mu}{T}+\frac{\mu^2}{2T^2}+\mathcal{O}(\frac{\mu^3}{T^3})\right
],
\end{eqnarray}
we obtain
\begin{eqnarray}\label{collision_a}
\frac{(\chi_{ij},\mathcal{C}\chi_{ij})^{(a)}}{e^4\ln{\frac{1}{e}}}&\approx&\int_0^\infty
dp\ \{f(p)[1-f(p)]+\bar f(p)[1-\bar f(p)]\}\nonumber\\
&&\times (1+\frac{3}{\pi^2}\ \frac{\mu^2}{T^2})\
\{p^2[\chi^e(p)']^2+6[\chi^e(p)]^2\}.
\end{eqnarray}

\subsection{Pair Production}
The pair production process is described by Fig.1(b) and its
reversed process. The typical matrix element for this process is:
\begin{equation}
|\mathcal{M}_{\gamma\gamma}^{ee}|^2=\frac{u}{t}+\frac{t}{u}\rightarrow\frac{2u}{t}=8e^4\frac{2pk}{q^2}(1-\cos\phi).
\end{equation}
Adding up the distribution functions contributions, the
equation(\ref{Boltamann3}) is recast into
\begin{eqnarray}\label{Boltamann_b1}
(\chi_{ij},\mathcal{C}\chi_{ij})^{(b)}=&&\frac{16e^4}{(4\pi)^6}\int_0^\infty
dq\int_{-q}^q d\omega\int_0^\infty dp \int_0^\infty dk
\int_0^{2\pi}d\phi(1-\cos\phi) \frac{2pk}{q^2}\nonumber\\[0.3cm]
&&\{ f(p)\bar f(k)[1+b(p)][1+b(k)]+\bar f(p)
f(k)[1+b(p)][1+b(k)]\nonumber\\[0.3cm]
&& +b(p)b(k)[1-\bar f(p)][1-f(k)]+b(p)b(k)[1- f(p)][1-\bar
f(k)]\}\nonumber\\[0.3cm]
&& \ \ \ \ \ \times
[\chi_{ij}^a(\mathbf{p})+\chi_{ij}^b(\mathbf{k})-\chi_{ij}^c(\mathbf{p'})-\chi_{ij}^d(\mathbf{k'})]^2.
\end{eqnarray}
Expanding the $\chi$-function term and ignoring the momenta
difference between the incoming and outgoing particles we get
\begin{eqnarray}
&&[\chi_{ij}^e(\mathbf{p})+\chi_{ij}^e(\mathbf{k})-\chi_{ij}^\gamma(\mathbf{p'})-\chi_{ij}^\gamma(\mathbf{k'})]^2\nonumber\\[0.3cm]
&&\ \ \ \ \ \approx I^2_{ij}(\hat p)[\chi^e(p)-\chi^\gamma(p)]^2+I^2_{ij}(\hat k)[\chi^e(k)-\chi^\gamma(k)]^2\nonumber\\[0.3cm]
&&\ \ \ \ \ \ \ \ \ +2 I_{ij}(\hat p)\cdot I_{ij}(\hat
k)[\chi^e(p)-\chi^\gamma(p)]\ [\chi^e(k)-\chi^\gamma(k)].
\end{eqnarray}
And noticing $I^2_{ij}(\hat p)=1$ and
\begin{equation}
I_{ij}(\hat p)\cdot I_{ij}(\hat
k)=\frac{1}{2}(3\cos^2\theta_{pk}-1)=P_2(\cos\theta_{pk})
\end{equation}
where $P_2(\cos\theta_{pk})$ is the second Legendre polynomial,
one can check the cross term vanishes when integrating over
$d\phi$. We carry out the k-integration by expanding the integrand
in terms of small $\mu/T$ and find,
\begin{eqnarray}
\int_0^\infty dk \ k \ \bar f(k)[1+b(k)]=T^2\left [\frac{\pi^2}{8}-0.963\frac{\mu}{T}+0.298\frac{\mu^2}{T^2}+\mathcal{O}(\frac{\mu^3}{T^3})\right ],&&\\[0.3cm]
\int_0^\infty dk \ k\ f(k)[1+b(k)]=T^2\left [\frac{\pi^2}{8}+0.963\frac{\mu}{T}+0.298\frac{\mu^2}{T^2}+\mathcal{O}(\frac{\mu^3}{T^3})\right ],&&\\[0.3cm]
\int_0^\infty dk \ k \ b(k)[1-f(k)]=T^2\left [\frac{\pi^2}{8}-0.270\frac{\mu}{T}-0.048\frac{\mu^2}{T^2}+\mathcal{O}(\frac{\mu^3}{T^3})\right ],&&\\[0.3cm]
\int_0^\infty dk \ k \ b(k)[1-\bar f(k)]=T^2\left
[\frac{\pi^2}{8}+0.270\frac{\mu}{T}-0.048\frac{\mu^2}{T^2}+\mathcal{O}(\frac{\mu^3}{T^3})\right
].&&
\end{eqnarray}
And then the equation(\ref{Boltamann_b1}) becomes
\begin{eqnarray}\label{Boltamann_b2}
(\chi_{ij},\mathcal{C}\chi_{ij})^{(b)}=&&\frac{\beta
e^4\ln\frac{1}{e}}{2^4\pi^5}\int_0^{\infty}dp\ p\ \left \{
f(p)[1+b(p)]
\left(\frac{\pi^2}{8}-0.963\frac{\mu}{T}+0.298\frac{\mu^2}{T^2}\right)
\right. \nonumber \\[0.3cm]
&&+\bar f(p)[1+b(p)]
\left(\frac{\pi^2}{8}+0.963\frac{\mu}{T}+0.298\frac{\mu^2}{T^2}\right)\nonumber \\[0.3cm]
&&+b(p)[1-\bar
f(p)]\left(\frac{\pi^2}{8}-0.27\frac{\mu}{T}-0.048\frac{\mu^2}{T^2}\right)\nonumber \\[0.3cm]
&&+b(p)[1-
f(p)]\left(\frac{\pi^2}{8}+0.27\frac{\mu}{T}-0.048\frac{\mu^2}{T^2}\right)\nonumber \\[0.3cm]
&&\times[\chi^e(p)-\chi^\gamma(p)]^2.
\end{eqnarray}

\subsection{Compton Scattering}
The Compton scattering process involves both electron and position
contributions. The matrix element for this process is
\begin{equation}
|\mathcal{M}_{e\gamma}^{e\gamma}|^2=
-8e^4\frac{s}{u}=8e^4\frac{2pk}{q^2}(1-\cos\phi).
\end{equation}
The distribution functions for this process is
\begin{equation}
f(p)b(k)[1-f(k)][1+b(p)]+\bar f(p)b(k)[1-\bar f(k)][1+b(p)].
\end{equation}
The $\chi$-function terms becomes
\begin{eqnarray}
&&\big[\chi_{ij}^e(\mathbf{p})+\chi_{ij}^\gamma(\mathbf{k})-\chi_{ij}^e(\mathbf{k})-\chi_{ij}^\gamma(\mathbf{p})\big]^2\nonumber\\[0.3cm]
&&\ \ \ \ \ \longrightarrow
\big[\chi^e(p)-\chi^\gamma(p)\big]^2+\big[\chi^e(k)-\chi^\gamma(k)\big]^2.
\end{eqnarray}

After finishing the integration over $dk$ we can recast the
equation(\ref{Boltamann3}) into
\begin{eqnarray}\label{Boltamann_c}
(\chi_{ij},\mathcal{C}\chi_{ij})^{(c)}=\frac{\beta
e^4\ln\frac{1}{e}}{2^3\pi^5}\int_0^{\infty}dp\ p\ \left \{
f(p)[1+b(p)]
\left(\frac{\pi^2}{8}-0.270\frac{\mu}{T}-0.048\frac{\mu^2}{T^2}\right)
\right.&& \nonumber \\[0.3cm]
+\left.\bar f(p)[1+b(p)]
\left(\frac{\pi^2}{8}-0.270\frac{\mu}{T}-0.048\frac{\mu^2}{T^2}\right)\right\}\big[\chi_{ij}^e(p)-\chi_{ij}^\gamma(p)\big]^2.\
\ \ &&
\end{eqnarray}

\section{Variational Method}
As far as shear viscosity is concerned, two species of particles
are involved and  $\chi(p)$ must take two components
\begin{equation}
\chi(p)= \left(
\begin{array}{c}
\chi^e(p) \\
\chi^\gamma(p)
\end{array}
\right).
\end{equation}
Accordingly the collision operator $\mathcal{C}$ is a $2\times 2$
matrix. The left hand side of Boltzmann equation(\ref{leftside})
reads
\begin{eqnarray}
(\chi_{ij},S_{ij})&=&-\frac{\beta^2}{\pi^2}\int^\infty_0 dp\ p^3\
\Big\{b(p)[1+b(p)]\chi^\gamma(p)\nonumber\\[0.3cm]
 &&+f(p)[1-f(p)]\chi^e(p)+\bar f(p)[1-\bar
f(p)]\chi^e(p)\Big\}.
\end{eqnarray}

Since we have already obtained all the collision terms in
Boltzmann equation, we are going to solve the equation
\begin{equation}
(\chi_{ij},S_{ij})=(\chi_{ij},\mathcal{C} \chi_{ij})
\end{equation}
to get the shear viscosity by variating the ansatz $\chi_{ij}$ to
reach its maximum value. We are not going to argue much about the
accuracy of this method in this paper, because Arnold {\itshape{et
al}}\cite{Arnold1} have compared it with the exact results at high
temperature but zero chemical potential environment. And we will
see the ansatz is the function only in terms of the momentum and
 the thermal variables, thereby we can safely use
the same ansatz form in the small $\mu$.

 Before we choose the exact ansatz of $\chi^\gamma$ and
$\chi^e$, we prefer to demonstrate the scheme of this variational
calculus. For simplicity all the subscripts and momentum
dependence of each function and operator are dropped out, and the
Boltzmann equation becomes
\begin{equation}\label{Boltamann_s}
(\chi,S)=(\chi,\mathcal{C}\chi).
\end{equation}
Expanding the $\chi $-function in a finite basis set
\begin{equation}
\chi=\sum\limits_{m=1}^N a_m\phi_m=\vec a\cdot\vec \phi
\end{equation}
one finds the equation(\ref{Boltamann_s}) becomes
\begin{equation}
\sum\limits_ma_m(\phi_m,S)= \sum\limits_{mn}a_m
a_n(\phi_m,\mathcal{C}\phi_m).
\end{equation}
Redefining $S$ and $\mathcal{C}$ in the {$\phi_m$} basis set, one
finds
\begin{equation}
\sum\limits_m a_m\tilde S_m=\sum\limits_{mn}a_m\
\tilde{\mathcal{C}}\ a_n
\end{equation}
with $\tilde S \equiv (\phi_m,S)$ and
$\tilde{\mathcal{C}}=(\phi_m,\mathcal{C}\phi_n)$. It's a trivial
exercise to give $\vec a=\tilde{\mathcal{C}}^{-1} \tilde S$ and
\begin{equation}\label{viscosity}
\eta=\frac{2}{15}Q_{max}=\frac{1}{15}\vec a\cdot\tilde
S=\frac{1}{15}\tilde{S^\top} \tilde{\mathcal{C}}^{-1}\tilde{S}.
\end{equation}

For the real two-component $\chi$-function one can expand it in
the same finite basis set
\begin{equation}
\chi^\gamma(p)=\sum\limits_{m=1}^N a_m \phi_m(p),\ \ \
\chi^e(p)=\sum\limits_{m=1}^N a_{N+m} \phi_m(p),
\end{equation}
where $\{a_1,a_2,\cdots,a_{2N}\}$ are the independent variational
parameters, and adopt the one function of the set with natural
ansatz $\phi(p)=p^2$,
\begin{equation}
\chi(p)= \left(
\begin{array}{c}
a_1 \phi(p)\\
a_2 \phi(p)
\end{array}
\right) =\left(
\begin{array}{c}
a_1 \\
a_2
\end{array}
\right) p^2 .
\end{equation}
By using this form of ansatz and neglecting the higher order than
$\mu^2/T^2$, one can evaluate
\begin{eqnarray}\label{leftside_final}
\tilde S&=&-\frac{\beta^2}{\pi^2}\int^\infty_0 dp\ p^3 \left(
   \begin{array}{c}
   b(p)[1+b(p)]\\[0.2cm]
   f(p)[1-f(p)]+\bar f(p)[1-\bar
    f(p)]
\end{array}
\right) p^2\nonumber\\[0.3cm]
&=&-\frac{120\xi(5)T^4}{\pi^2}
\left(
\begin{array}{c}
1\\[0.3cm]
\frac{15}{8}(1+0.869\frac{\mu^2}{T^2})
\end{array}
\right)
\end{eqnarray}
by expanding the fermion distribution function in term of $\mu/T$
and neglecting the higher order of $\mu^2/T^2$.

The collision term $\mathcal{\tilde{C}}$ can be obtained likewise
by combining equations (\ref{collision_a})(\ref{Boltamann_b2}) and
(\ref{Boltamann_c})
\begin{eqnarray}\label{collision_total}
\mathcal{\tilde{C}}&=&\frac{\pi T^5e^4\ln e^{-1}}{9}\left[ \left(
\begin{array}{cc}
0&0\\[0.3cm]
0&\frac{7}{4}(1+0.738\frac{\mu^2}{T^2})
\end{array}\right)
+\frac{9\pi^2}{128}(1+0.443\frac{\mu^2}{T^2})\left(
\begin{array}{cc}
1&-1\\[0.3cm]
-1&1
\end{array}
\right ) \right]\nonumber\\
\end{eqnarray}

Inserting equation (\ref{leftside_final}) and
(\ref{collision_total}) into equation (\ref{viscosity}) we obtain
the shear viscous coefficient for QED plasma
\begin{equation}\label{result}
\eta_{QED}=187.13\frac{T^3}{e^4\ln
e^{-1}}\left(1+0.13\frac{\mu^2}{T^2}+\mathcal{O}(\frac{\mu^4}{T^4})\right),
\end{equation}
which recovers the result of Ref.\cite{Arnold1} at $\mu=0$ and has
similar structure as that from relaxation time
approximation\cite{Defu,Thoma}
\section{Discussion and Outlook}

So far we have obtained the shear viscosity of QED plasma at
finite temperature and density in the leading-log order. The
chemical potential modifies the result in pure temperature case by
a small factor of $\mu^2/T^2$, which ensures the modification
factor is irrelevant to the sign of net charge of plasma due to
the symmetry. In addition, the sign in front of the modification
factor is positive, which indicates that the chemical potential
increases the shear viscosity of the plasma. Although we obtain
this result in small $\mu$ limit, the tendency keeps unchanged in
the whole region of $\mu<T$.

In the thermal field theory, we can also obtain such kind of
result like equation(\ref{result}) by replacing the damping rate
by the transport damping rate\cite{Liu} in the boson-exchange
case. The reason for this replacement is clear when one looks into
the kinetic theory: the extra $q^2$ coming from the
$\chi$-function in $ee\rightarrow ee$ scattering softens the
quadratic divergence into a logarithmical one. This extra small
$q^2$ is appeared only in boson-exchange process and is the origin
of extra $\sin^2\frac{\theta}{2}$ in the transport damping rate.
Carrington, Defu and Kobes also pointed out\cite{Carrington2},
these $\chi$-terms can be explained as an infinite series of
resummed ladder diagrams. These facts imply that the one-loop
calculation with usual interaction rate is not complete even in
the amplitude of order. But the replacement of transport rate
improves the calculation and makes the results reliable.

We have calculated the viscosity of plasma involving only
$2\rightarrow 2$ processes to leading logarithm. But the inelastic
scatterings and interference effects might be important if we go
beyond the leading-log and obtain the complete leading order
contribution. Furthermore, to explain the near-perfect property of
QGP, one need to treat the strong coupling system. In this case we
have to use Kubo formula and calculate the correlation functions
of relevant currents.\\[.5cm]

\centerline{\bf Acknowledgement}
 This work is partly supported by
the National Natural Science Foundation of China under project
Nos. 90303007 and 10135030 and the Ministry of Education of China
with Project No. 704035.

\end{document}